\documentclass[a4paper, 11pt, onecolumn]{article}

\usepackage{fullpage}
\usepackage[english]{babel}
\usepackage[pdftex]{color}

\usepackage{harvard}
\usepackage{amsmath}                 
\usepackage{graphicx}
\usepackage{url}     
\usepackage[utf8]{inputenc}
\usepackage{color}
\usepackage{soul}

\usepackage{setspace}

\RequirePackage[left=0.95in,right=0.9in,top=1in,bottom=1in]{geometry}

\begin{document}

\title{Learning to Rank Academic Experts in the DBLP Dataset}

\author{Catarina Moreira\\ \small{catarina.p.moreira@ist.utl.pt}\\
\and
P\'{a}vel Calado\\ \small{pavel.calado@ist.utl.pt}\\ 
\and
Bruno Martins\\ \small{bruno.g.martins@ist.utl.pt}
\and
\\Instituto Superior T\'{e}cnico, INESC-ID\\ Av. Professor Cavaco Silva, 2744-016 Porto Salvo, Portugal\\  
\\ \small The original publication is available at: Expert Systems, Wiley Online Library\\
 \small \text{\url{http://onlinelibrary.wiley.com/doi/10.1111/exsy.12062/abstract}}
}

\date{}

\maketitle 

\doublespace

\begin{abstract} 

Expert finding is an information retrieval task that is concerned with the search for the most knowledgeable people with respect to a specific topic, and the search is based on documents that describe people's activities. The task involves taking a user query as input and returning a list of people who are sorted by their level of expertise with respect to the user query. Despite recent interest in the area, the current state-of-the-art techniques lack in principled approaches for optimally combining different sources of evidence. This article proposes two frameworks for combining multiple estimators of expertise. These estimators are derived from textual contents, from graph-structure of the citation patterns for the community of experts, and from profile information about the experts. More specifically, this article explores the use of supervised learning to rank methods, as well as rank aggregation approaches, for combing all of the estimators of expertise. Several supervised learning algorithms, which are representative of the pointwise, pairwise and listwise approaches, were tested, and various state-of-the-art data fusion techniques were also explored for the rank aggregation framework. Experiments that were performed on a dataset of academic publications from the Computer Science domain attest the adequacy of the proposed approaches.
\\
\end{abstract} 

\section{Introduction}

The search for people who are knowledgeable about specific topic areas and who are within the scope of specific user communities, with basis on documents that describe people's activities, is an information retrieval problem that has been receiving increasing attention~\cite{Pavel09Search}. Usually referred to as \emph{expert finding}, this task involves taking a short user query as input and returning a list of people who are sorted by their level of expertise in what concerns the query topic. When looking for experts in an enterprise environment, the documents that describe people's activities include reports, web pages, and manuals. When looking for academic experts, these documents are the individual's publications.

Expert finding has been gaining attention not only in the scientific community but also in enterprises. Companies are interested in expert finding systems to save time and money. For example, when a problem arises in an on-going project, an expert with some specific knowledge must be found with some urgency. In large companies, going through all of the documentation and looking for people who might have the skills that are needed to solve the problem is time consuming. Thus, often enterprises engage external personnel to solve their issues, when the solution to their problem could be in an employee from their own ranks. The TREC enterprise track dataset from the 2008 edition simulates this scenario. It provides documents from an Australian company. The goal is for researchers to test their algorithms for finding the most knowledgeable people for some query topics. In an academic scenario, expert finding systems are also very useful for researchers and for students who are looking for the best advisor for their work.

Several effective approaches for finding experts have been proposed in the literature; these approaches explore different retrieval models and different sources of evidence for estimating expertise. However, the current state-of-the-art techniques still lack in principled approaches for combining the multiple sources of evidence that can be used to estimate expertise. In traditional information-retrieval tasks, such as ad hoc retrieval, there has been an increasing interest in the use of machine learning methods for building retrieval formulas that are capable of estimating relevance for query-document pairs. This approach is commonly referred to as Learning to Rank for Information Retrieval (L2R4IR)~\cite{Liu09Learning}. The general idea behind L2R4IR approaches is to use hand-labelled data (e.g., document collections that contain relevance judgments for specific sets of queries, or information regarding user-clicks that are aggregated over query logs) to train ranking models and, in this way, use data to combine the different estimators of relevance in an optimal way. Thus far, although many different approaches have been proposed in the expert finding literature, few previous studies have specifically addressed the use of learning to rank in the development of approaches for the task of expert finding. 

The combination of multiple sources of evidence has also received a substantial amount of interest in traditional search engines. The problem of combining various ranked lists for the same set of documents, to obtain a more accurate and more reliable ordering, can be defined as Rank Aggregation~\cite{Dwork01rankaggr}. Data fusion techniques include methods that are used to combine these different rankings.~\citeasnoun{Montague02condorcet} have experimented with data fusion techniques in the domain of search engines, where they concluded that data fusion can provide significant advantages. Different retrieval methods often return very different irrelevant documents, although they return the same relevant ones. Thus, rank aggregation can provide a more reliable performance than individual retrieval methods.	

This article explores the use of learning to rank methods or, alternatively, the use of rank aggregation algorithms in the expert finding task, specifically combining a large pool of estimators for expertise. We build on a preliminary study by~\citeasnoun{moreira11epia} that addresses the same problem, adding a larger set of experiments. We have evaluated this work on an academic publication dataset from the Computer Science domain. 

The main contributions of this work can be summarised as follows:

\begin{itemize}
	\item A Set of Features to Estimate Expertise. In this study, we defined a set of features that are based on three different sources of evidence, namely textual similarity features, author profile information features and features based on citation graphs. The textual features use traditional information retrieval techniques, which measure term co-occurrences between the query topics and the documents that are associated with a candidate expert. The profile information features measure the total publication record of a candidate throughout his career, under the assumption that true expert candidates are more productive. Finally, the features that are based on citation graphs capture the authority of candidate experts from the attention that others give to their work. Although we might say that all of the features correspond to statistical values that are often used in the literature in different areas, to the best of our knowledge, the present study is the first to use them for the task of discovering experts. The only exceptions are the numbers of publications and citation features, which have been used in the previous work of~\citeasnoun{Yang09bole}. In addition, we applied a new set of features that has never been used before in Information Retrieval or in expert finding. Inspired by works in Scientometrics, we tested a set of academic indexes, which are commonly used to estimate the impact of the author's publications on the scientific community: h-index, g-index, e-index, contemporary h-index, trend h-index, a-index and the individual h-index. In terms of the features that we proposed, the use of such indexes is our main contribution.
	
	\item A Supervised Learning to Rank Approach for Expert Finding. Our experiments explored the use of learning to rank methods in the expert finding task, specifically combining a large pool of estimators of expertise, and we used the hypothesis that learning to rank approaches provide a significant improvement over the current state-of-the-art methods. To combine these multiple estimators, we performed experiments with state-of-the-art algorithms from the pointwise, pairwise and listwise learning to rank approaches. There are a couple of approaches in the literature that propose a learning to rank framework for the problem of expert finding. One approach is concerned with expertise retrieval in enterprises~\cite{Macdonald11aggr}, and the other approach is based on finding academic experts~\cite{Yang09bole}, which is similar to the method used in our paper. Our main concern with both of these approaches is that they lack a detailed description of how the learning to rank framework works, and they formulate experiments by using only two algorithms (AdaRank and Metzler's Automatic Feature Selection algorithm in~\citeasnoun{Macdonald11aggr} and SVMrank in~\citeasnoun{Yang09bole}). In our paper, we provide a more thought-out description of how the learning to rank framework works, and we also provide a set of algorithms that range from the various classes of learning to rank solutions, with which we made comparative experiments.
	
	\item A Rank Aggregation Approach for Expert Finding. Our experiments also tested the hypothesis that rank aggregation methods, which are based on data fusion techniques, can provide significant advantages over the representative generative probabilistic models that are proposed in the expert finding literature. We use existing state-of-the-art algorithms to build a single expert finding model that enables the combination of a large pool of expertise estimates.
	
\end{itemize}

The remainder of this article is organised as follows: Section~\ref{sec:relatedwork} presents the main concepts and related work. Section~\ref{sec:l2r} presents the learning to rank approaches that are used in our experiments. Section~\ref{sec:rankaggregation} details the rank aggregation framework as well as the data fusion techniques that are used in our experiments. Section~\ref{features} introduces the multiple features that we use to estimate the expertise. Section~\ref{sec:validation} describes how the system was evaluated, detailing the datasets that are used in our experiments as well as the obtained results. Finally, Section~\ref{sec:conclusions} presents our conclusions and points to directions for future work in this area.

\section{Concepts and Related Work}~\label{sec:relatedwork}

Previous publications have surveyed the most important concepts and representative previous studies in the expert finding task~\cite{Pavel09Search,Macdonald08Voting}. Two of the most popular and well-performing methods are the candidate-based and the document-based approaches. In candidate-based approaches, the system gathers all of the textual information about a candidate and merges it into a single document (i.e., the profile document). The profile document is then ranked by determining the probability of the candidate given the query topics. Candidate-based approaches are also referred to as Model 1 in~\citeasnoun{Balog06FormalModels} and are referred to as query-independent methods according to~\citeasnoun{Petkova06expertfinding}. In document-based approaches, the system gathers all of the documents that contain the expertise topic terms that are included in the query. Then, the system uncovers which candidates are associated with each of those documents and determines their probability scores. The final ranking of a candidate is provided by summing all of the individual scores of the candidate in each document. Document-based approaches are also referred to as Model 2 in~\citeasnoun{Balog06FormalModels} or as query-dependent methods according to~\citeasnoun{Petkova06expertfinding}. Experimental results show that document-based approaches usually outperform candidate-based approaches~\cite{Balog06FormalModels}.

The first candidate-based approach was proposed by~\citeasnoun{Craswell01panoptic}, where the ranking of a candidate was computed through text similarity measures that were computed between the query topics and the candidate's profile document. ~\citeasnoun{Balog06FormalModels} formalised a general probabilistic framework for modelling the expert finding task, which used language models to rank the candidates. Language models apply probabilistic functions that rank documents based on the probability of the document model generating the query topic (within the document). This scenario occurs when the document contains a large number of occurrences of the query terms~\cite{Manning08IntroductionIR}.
\citeasnoun{Petkova06expertfinding}~presented a general approach for representing the knowledge of a candidate expert as a mixture of language models from associated documents. Later, ~\citeasnoun{Balog09LanguageModels} and~\citeasnoun{Petkova07namedentity} introduced the idea of dependency between candidates and query topics by including a surrounding window to weight the strength of the associations between mentions to candidates in the text of the documents and query topics. The surrounding window measures the proximity in which two words occur in the text. For expert finding, this proximity plays an important role because when the query topics appear next to a candidate's name, there is a high probability that this query topic is associated with that candidate.

Many different authors have also proposed sophisticated probabilistic retrieval models that are based on the document-based approaches~\cite{Balog06FormalModels,Petkova07namedentity,Pavel09Search}. For example, ~\citeasnoun{Cao05trec} proposed a two-stage language model that combines document relevance and co-occurrence between experts and query terms. ~\citeasnoun{Fang07expertsearch} derived a generative probabilistic model from the probabilistic ranking principle and extended it with query expansion and non-uniform candidate priors. ~\citeasnoun{zhu07trec} proposed a multiple-window-based approach for integrating multiple levels of associations between experts and query topics in expert finding. Later, ~\citeasnoun{Zhu08Modeling} proposed a unified language model that integrates many document features. 

In addition to the candidate-based and document-based approaches, other methods have also been proposed in the expert finding literature. For example, ~\citeasnoun{Macdonald08Voting} formalised a voting framework that was combined with data fusion techniques. Each candidate that was associated with documents that contained the query topics received a vote, and the ranking of each candidate was given by the aggregation of the votes of each document through data fusion techniques. ~\citeasnoun{deng11enhanced} proposed a query-sensitive AuthorRank model. These investigators modelled a co-authorship network and measured the weights of the connections between authors with the AuthorRank algorithm~\cite{liu05networks}. Because AuthorRank is query independent, the authors added probabilistic models to refine the algorithm to consider the query topics. ~\citeasnoun{serdyukov08mixtures} proposed the person-centric approach, which combines the ideas of the candidate and document-based approaches. Their system starts by retrieving the documents that contain the query topics, and then, it ranks the candidates by combining the probability of generation of the query by the candidate's language model.

Although these models are capable of employing different types of associations among query terms, documents and experts, they mostly ignore other important sources of evidence, such as the importance of individual documents or the citation patterns between candidate experts that are available from citation graphs. In this paper, we study two different principled approaches for combining a much larger set of estimates for expert finding, namely learning to rank and rank aggregation.

Another work that follows the paradigm of this paper belongs to ~\citeasnoun{Macdonald11aggr}, who proposed a learning to rank approach in which they created a feature generator that was composed of three components, namely, a document ranking model, a cutoff value to select the top documents according to the query topics and rank aggregation methods. Using those features, the authors made experiments with the AdaRank listwise learning to rank algorithm, which outperformed all of the generative probabilistic methods that were proposed in the literature.

\citeasnoun{Fang07expertsearch}~applied the probabilistic ranking principle to develop a general framework from which the candidate-based and document-based models for expert finding could be derived. They also showed how query expansion techniques, such as the association of different weights to each candidate representation and the topic expansion to give more textual information that is related to the original query, can improve the performance of the models by using the framework.

In the Scientometrics community, the evaluation of the scientific output of a scientist has also attracted significant interest, due to the importance of obtaining unbiased and fair criteria. Most of the existing methods are based on metrics such as the total number of authored papers or the total number of citations~\cite{Sidiropoulos05citation,Sidiropoulos06generalized}. Simple and elegant indexes, such as the Hirsch index criteria, calculate how broad the research work of a scientist is, accounting for both productivity and impact. Graph centrality metrics inspired by PageRank, which are calculated over co-authorship graphs, have also been extensively used~\cite{liu05networks}. In the context of academic expert search systems, these metrics can easily be used as query-independent estimators of expertise, in much the same way that PageRank is used in the case of Web information retrieval systems.

A comprehensive survey about expertise retrieval and the different techniques proposed in the literature can be found in~\cite{Balog12}.

\subsection{Literature Regarding Learning to Rank Algorithms}

For combining the multiple sources of expertise, we propose to use previous work concerning the subject of L2R4IR.~\citeasnoun{Liu09Learning} presented a notable survey on the subject, categorising the previously proposed supervised L2R4IR algorithms into three groups, according to their input representation and optimisation objectives:

\begin{itemize}
\item {\bf Pointwise approach} - L2R4IR is seen as either a regression or a classification problem. Given the feature vectors of each single document from the data for the input space, the relevance degree of each of those individual documents is predicted with either a regression or a classification model. The relevance scores can then be used to sort the documents to produce the final ranked list of results. Several different pointwise methods have been proposed in the literature, including the Additive Groves algorithm by~\citeasnoun{Sorokina07groves}, RankClass~\cite{Ji11}, the algorithm proposed by~\citeasnoun{Adali07} and random model trees~\cite{Pfahringer11}.

\item {\bf Pairwise approach} - L2R4IR is seen as a binary classification problem for document pairs because the relevance degree can be regarded as a binary value that tells which document order is better for a given pair of documents. Given the feature vectors of pairs of documents from the data for the input space, the relevance degree of each of those documents can be predicted with scoring functions that attempt to minimise the average number of misclassified pairs. Several different pairwise methods have been proposed, including SVMrank~\cite{Joachims06rank}, RankNet~\cite{burges05ranking}, RankBoost~\cite{freund03ranking} and P-Norm Push~\cite{Ertekin11}.

\item {\bf Listwise approach} - L2R4IR is addressed in a way that accounts for an entire set of documents that are associated with a query, taking each document as an instance. These methods train a ranking function through the minimisation of a listwise loss function that is defined on the predicted list and the ground truth list. Given feature vectors of a list of documents of the data for the input space, the relevance degree of each of those documents can be predicted with scoring functions that attempt to directly optimise the value of a specific information-retrieval evaluation metric, which is averaged over all of the queries in the training data~\cite{Liu09Learning}. Several different listwise methods have also been proposed, including SVMmap~\cite{Yue07Support}, AdaRank~\cite{Xu08directlyoptimizing,xu07ranking}, Coordinate Ascent~\cite{Metzler07linearfeature-based} and P-Classification~\cite{Ertekin11}.
\end{itemize}

There are also some works that have extended the learning to rank approach to a relational learning to rank framework~\cite{qin08learning}. In this new learning task, the ranking model accounts for not only the features in the documents but also the relationship information between the documents. The main difference between traditional learning to rank tasks and this new approach is that, in relational learning to rank, the ranking function is not solely concerned with the optimisation of a specific bound for the retrieval task. Instead, it attempts to develop relationships between the documents that are retrieved. For example, if we have two very similar documents and the learning function only considers one of them to be relevant, then through a relationship, the other document will be given a ranking that is similar  to the first document.

\subsection{Literature Regarding Data Fusion Algorithms}

For combining the different sources of expertise evidence, we also rely on previous studies that have addressed the problem of ranking search results through a rank aggregation framework, which are often based on data fusion methods that take their inspiration from voting protocols proposed in the area of social sciences.~\citeasnoun{Riker88} suggested a classification to distinguish the different existing data fusion algorithms into two categories, namely the positional methods and the majoritarian algorithms. Later,~\citeasnoun{Fox94Combination} proposed score aggregation methods that are specifically designed for information retrieval.

The positional methods are characterised by the computation of a candidate's score based on the position that the candidate occupies in the ranked lists given by each voter. If the candidate falls into the top position of the ranked list, then he receives a maximum score. If the candidate falls into the end of the list, then his score is a minimum score. The most representative positional algorithms are the Borda Count~\cite{borda81} and the Reciprocal Rank~\cite{Voorches99trec} fusion methods.

The majoritarian algorithms are characterised by a series of pairwise comparisons between candidates. The candidates are scored according to the number of times that they win another candidate in a pairwise comparison. The most representative majoritarian algorithm is most likely the Condorcet Fusion method proposed by~\citeasnoun{Montague02condorcet}. However, there have been other proposals that are based on Markov Chain Models~\cite{Dwork01rankaggr}.

Finally, score aggregation methods determine the highest ranked candidate by combining the ranking scores from all of the input rankings.~\citeasnoun{Fox94Combination} proposed the CombSUM and CombMNZ methods, which have been used frequently in IR experiments.
In this article, we performed experiments with representative supervised learning to rank algorithms from the pointwise, pairwise and listwise approaches, as well as with representative state-of-the-art data fusion algorithms from the positional, majoritarian and score aggregation approaches. Sections~\ref{sec:l2r} and~\ref{sec:rankaggregation} detail, respectively, the learning to rank and rank aggregation approaches.

\section{The Framework for Learning to Rank Experts}~\label{sec:l2r}

One of the research questions that motivates this work is concerned with the possibility of learning to rank approaches being effectively used in the context of expert search tasks to combine different estimators of expertise in a principled way, to improve on the current state-of-the art methods.
 
\begin{figure}%
	\centering
		\includegraphics[scale = 0.5]{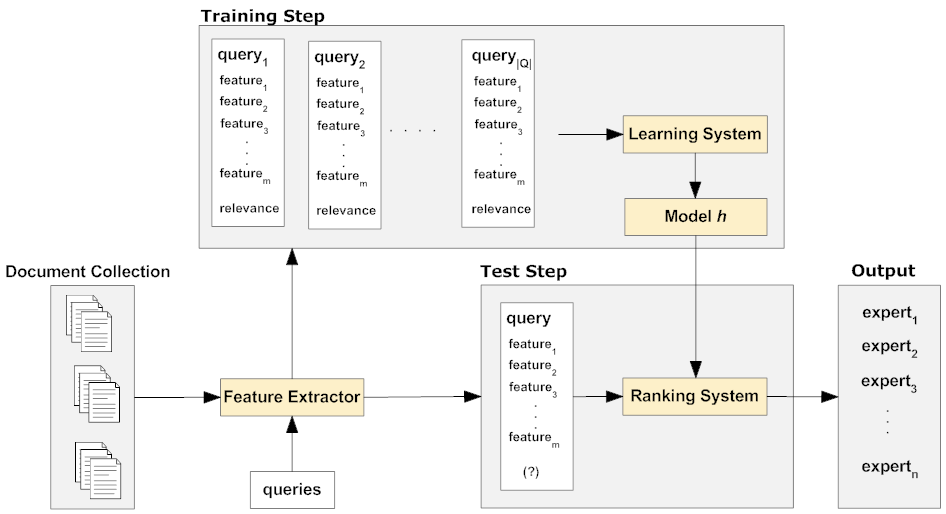}%
    \caption{The learning to rank framework for expert finding}
     \label{fig:expertl2r}
\end{figure}
 
The expert finding problem can be formalised as follows. Given a set of queries $Q = \{q_1,\ldots, q_{m}\}$ and a collection of experts $E = \{e_1,\ldots, e_{n}\}$, each of which is associated with specific documents that describe his topics of expertise, a training corpus for learning to rank is created as a set of query-expert pairs, $(q_i, e_j) \in Q \times E$, upon which a relevance judgment that indicates the match between $q_i$ and $e_j$ is assigned by a labeller. This relevance judgment is a binary label that indicates whether the expert $e_j$ is relevant to the query topic $q_i$ or not. For each instance $(q_i, e_j)$, a feature extractor produces a vector of features that contains statistical values that are related to $q_i$ and $e_j$. The features can range from classical IR estimators computed from the documents associated with the experts (e.g., term frequency, inverse document frequency) to link-based features that are computed from networks that encode relations between the experts (e.g., PageRank). These features are detailed in Section~\ref{features} of this article. The inputs to the learning algorithm comprise training instances, their feature vectors, and the corresponding relevance judgments. The output is a ranking function, $h$, which produces a ranking score for each candidate expert $e_j$ in such a way that, when sorting experts for a given query according to these scores, the more relevant experts appear on the top of the ranked list.

During the training process, the learning algorithm attempts to learn a ranking function that is capable of sorting experts in the following ways: for the listwise approach, it optimises a specific retrieval performance measure; for the pairwise approach, it attempts to minimise the number of misclassifications between the expert pairs; and for the pointwise approach, it attempts to directly predict the relevance score. In the test phase, the learned ranking function is applied to determine the relevance between each expert $e_j$ in $E$ toward a new query $q_i$. Figure~\ref{fig:expertl2r} shows a general illustration of the learning to rank framework that is used in this work. The experiments reported in this paper compared many different learning to rank algorithms, which required manual tuning of different parameters. In the next sections, we describe how we adjusted these parameters, and we also detail how the different methods work.

\subsection{Different Families of Learning to Rank Algorithms}

In this section, we describe the algorithms that were used in our experiments, categorising them by the learning method that was used.

\subsubsection{ Algorithms based on Boosting Theory }

Boosting theory is a supervised machine learning approach that iteratively attempts to improve a candidate solution. Boosting theory uses the concepts of weak and strong learners. Weak learners are classifiers that are slightly correlated with true classification. Strong learners are classifiers that are highly correlated with the true classification. The paradigm of boosting theory involves creating a single strong learner through the combination of a set of weak learners~\cite{freund03ranking}.

The AdaRank listwise method, proposed by~\citeasnoun{Xu08directlyoptimizing}, builds a ranking model through the formalism of boosting, attempting to optimise a specific information retrieval performance measure. The basic idea of AdaRank is to train one weak ranker at each round of iterations and to combine these weak rankers as the final ranking function. After each round, the experts are re-weighted by decreasing the weight of correctly ranked experts, based on a specific evaluation metric, and by increasing the weight of the experts that performed poorly for the same metric. The AdaRank algorithm receives as input the parameter $T$, which is the number of iterations that the algorithm will perform, and the parameter $E$, which corresponds to a specific information retrieval performance measure. 

The RankBoost pairwise method, proposed by~\citeasnoun{freund03ranking}, also builds a ranking model through the formalism of boosting, attempting to minimise the number of misclassified pairs of experts in a pairwise approach. The basic idea of RankBoost is to train one weak ranker at each round of iteration and to combine these weak rankers as the final ranking function. After each round, the expert pairs are re-weighted by decreasing the weight of correctly ranked pairs of experts and increasing the weight of wrongly ranked experts. The RankBoost algorithm receives as input the parameter $T$, which is the number of iterations that the algorithm will perform, and the parameter $\theta$, which is a threshold that corresponds to the number of candidates to be considered in the weak rankers. 

The Additive Groves pointwise method, introduced by~\citeasnoun{Sorokina07groves}, builds a ranking model through the formalism of regression trees, attempting to directly minimise the errors in relevance predictions over the training dataset. In this approach, a {\it grove} is an additive model that contains a small number of large trees. The ranking model of a grove is built upon the sum of the ranking models of each one of those trees. The basic idea of Additive Groves is to initialise a grove with a single small tree. Iteratively, the grove is gradually expanded by adding a new tree or by enlarging the existing trees of the model. The new trees in the grove are trained with the set of experts that were misclassified by the other previously trained trees. In addition, trees are discarded and retrained in turn until the overall predictions converge to a stable function. The goal of this algorithm is to find the simplest model that can make the most accurate predictions. The prediction of a grove is given by the sum of the predictions of the trees that are contained in it. 

One major problem in using regression trees is that these models will learn a function that fits the training data very well, but cannot make good predictions under unseen data. This phenomenon is known as overfitting, and these models tend to have high variances. Therefore, there is a high danger of regression trees overfitting the training data. The bagging procedure improves the performance of these models by reducing the variance. Thus, bagging avoids overfitting the training data.

The algorithm receives as input the parameter $N$, which is the number of trees in the grove, the parameter $\alpha$, which controls the size of each individual tree, and the parameter $b$, which is the number of bagging iterations, i.e., the number of additive models that are combined in the final ensemble. The publicly available version of this algorithm tuned these parameters automatically.

\subsubsection{Algorithms Based on Classification by N-Dimensional Hyperplanes}

Artificial Neural Networks (ANNs) are a machine learning approach that attempts to construct an N-dimensional decision boundary surface that separates data into positive examples and negative examples, through the simulation of some properties that occur in biological neural networks. They require the use of optimisation methods, such as gradient descent, to find a solution that minimises the number of misclassifications~\cite{haykin08}.

The RankNet pairwise method, proposed by~\citeasnoun{burges05ranking}, builds a ranking model through the formalism of Artificial Neural Networks, attempting to minimise the number of misclassified pairs of experts. The basic idea of RankNet is to use a multilayer neural network with a cost error entropy function. While a typical artificial neural network computes this cost by measuring the difference between the network's output values and the respective target values, RankNet computes the cost function by measuring the difference between a pair of network outputs. RankNet attempts to minimise the value of the cost function by adjusting each weight in the network according to the gradient of the cost function. This goal is accomplished through the use of the backpropagation algorithm. The RankNet algorithm receives as input the parameter \emph{epochs}, which is the number of iterations that are used in the process of providing the network with an input and updating the network's weights, and the parameter \emph{hiddenNodes}, which corresponds to the number of nodes that are in the network's hidden layer. If there are too few nodes, then we can underfit the data. On the other hand, if there are too many nodes, then we can overfit the data, and the resulting network will not generalise well. 

Support Vector Machines (SVMs) can also be defined as learning machines that construct an N-dimensional decision boundary surface that optimally separates data into positive examples and negative examples, by maximising the margin of separation between these examples. One major advantage is the computational complexity of an SVM, which does not depend on the dimensionality of the input space~\cite{haykin08}.

The SVMmap listwise method, introduced by~\citeasnoun{Yue07Support}, builds a ranking model through the formalism of structured Support Vector Machines~\cite{Tsochantaridis05Structured}, attempting to optimise the metric of Average Precision (see Section~\ref{sec:exp}). The basic idea of SVMmap is to minimise a loss function that measures the difference between the performance of a perfect ranking (i.e., when the Average Precision equals one) and the minimum performance of an incorrect ranking. The SVMmap algorithm receives as input the parameter $C$, which affects the trade-off between the model complexity and the proportion of non-separable samples. If $C$ is too large, then we have a high penalty for non-separable points and we could create many support vectors, which could lead to overfitting. If $C$ is too small, then we could have underfitting. In our experiments, we used SVMmap with a radial basis function kernel that also requires the manual tuning of the parameter $\gamma$, which determines the area of influence that the centre support vector has over the data space. 

The SVMrank pairwise method, introduced by~\citeasnoun{Joachims06rank}, builds a ranking model through the formalism of Support Vector Machines. The basic idea of SVMrank is to attempt to minimise the number of misclassified expert pairs in a pairwise setting. This goal is achieved by modifying the default support vector machine optimisation problem by constraining the optimisation problem to perform a minimisation of each pair of experts. This optimisation is performed over a set of training queries, their associated pairs of experts and the corresponding relevance judgment over each pair of experts (i.e., pairwise preferences that result from a conversion from the ordered relevance judgments over the query-expert pairs). SVMrank receives as input the parameter $C$, which affects the trade-off between the model complexity and the proportion of non-separable samples. In our experiments, we used a linear kernel.

\subsubsection{Algorithms based on Unconstrained Optimisations}

The Coordinate Ascent listwise method, proposed by~\citeasnoun{Metzler07linearfeature-based}, is an optimisation algorithm that is used in unconstrained optimisation problems and that builds a ranking model by directly maximising an information retrieval performance measure. The basic idea of Coordinate Ascent is to iteratively optimise a multivariate objective function by solving a series of one-dimensional searches. In each iteration, Coordinate Ascent randomly selects one feature to perform a search on while holding all of the other features. This way, in each iteration, the algorithm chooses the parameters that maximise the information retrieval performance measure. The Coordinate Ascent algorithm receives as input the parameter $rr$, which is the number of random restarts, and the parameter $T$, which corresponds to the number of iterations to perform in each one-dimensional space.

\subsection{Parameter Estimation}\label{sec:parameter_est}

The most naive approach to parameter search is the grid search method. In this approach, a grid is placed over the parameter space, and the data are evaluated at every grid intersection, returning the parameters that lead to maximum performance of the learning algorithm~\cite{Metzler07linearfeature-based}. However, a grid search has the problem of being unbounded because an infinite set of parameters is available for testing. To overcome this issue, the parameter search was restricted by using the boundaries that were suggested by~\cite{Hsu10lib} for the SVM-based algorithms ($C \in \{ 2^{-5}, 2^{-3}, ..., 2^{13}, 2^{15}\}$ and $\gamma \in \{ 2^{-5}, 2^{-3}, ..., 2^{13}, 2^{15} \}$). For the other approaches (AdaRank, Coordinate Ascent, RankBoost and RankNet), the grid search was stopped when the results that were obtained started to converge into a single value according to an information retrieval metric. The parameters were learned with a k-fold cross-validation method and were fitted to the training data. We collected the parameters that, on average, achieved the best results over all of the tested folds.

\section{The Framework for Aggregating Expert Rankings}~\label{sec:rankaggregation}

\begin{figure}[ht]
\centering
\includegraphics[width=\columnwidth]{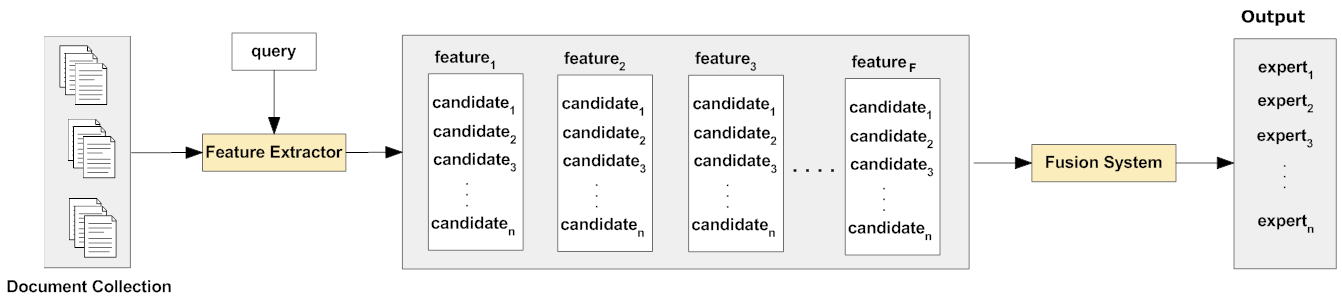}
\caption{A general rank aggregation framework for expert retrieval}
\label{fig:dataFusion}
\end{figure}

We also experimented with the use of rank aggregation frameworks for combining multiple sources of expertise evidence. Because data fusion can aggregate the rankings of several individual features for each candidate, it has the advantage of not reflecting the tendency of a single feature. Instead, it reflects the combination of all of them, resulting in a more reliable and accurate ranking system. The general rank aggregation framework that is proposed for expert finding is illustrated in Figure~\ref{fig:dataFusion}.

In this framework, we are given a set of queries $Q=\{q_1, q_2, ..., q_{m}\}$ and a collection of candidate experts $E = \{e_1, e_2,..., e_{n}\}$, each of which is associated with specific documents that describe the candidate's topics of expertise. For each instance $(q_i, e_j) \in Q \times E$, a feature extractor produces a set of ranked lists according to the match between $q_{i}$ and $e_j$. These features are detailed in Section~\ref{features} of this work. A data fusion algorithm is then applied to combine the various ranked lists that are computed by each of the features. The inputs of a rank aggregation algorithm comprise a set of queries and the data fusion technique to be applied. The output produces a ranking score that results from the aggregation of multiple features. The relevance of each expert $e_j$ towards the query $q_i$ is determined through this aggregated score.

The score aggregation data fusion techniques that are used in our experiments required normalised scores for the different features. To perform this normalisation, we applied the Min-Max Normalisation procedure, which is given by Equation~\ref{eq:norm}.

\begin{equation}
\text{NormalisedValue = }\frac{Value - minValue}{maxValue - minValue}
\label{eq:norm}
\end{equation}

The CombSUM, CombMNZ and CombANZ approaches, which were introduced by~\citeasnoun{Fox94Combination}, are three examples of rank aggregation algorithms.

In this article, we performed experiments with representative data fusion algorithms from the information retrieval literature, namely CombSUM, CombMNZ, CombANZ, Borda Fuse, Reciprocal Rank Fuse and Condorcet Fusion. 

The {\bf CombSUM} score of an expert $e$ for a given query $q$ is the sum of the normalised scores received by the expert in each individual ranking feature and is given by Equation~\ref{eq:CombSUM}.

\begin{equation}
\text{CombSUM(e,q) = }\sum_{j = 1} ^k score_{j}(e, q)
\label{eq:CombSUM}
\end{equation}

Similarly, the {\bf CombMNZ} score of an expert $e$ for a given query $q$ is defined by Equation~\ref{eq:CombMNZ}, where $r_e$ is the number of ranking features that contribute to the retrieval of the candidate, by having a score that is larger than zero.

\begin{equation}
\text{CombMNZ(e,q) = } \text{CombSUM(e, q)}\times r_e
\label{eq:CombMNZ}
\end{equation}

The {\bf CombANZ} score of an expert $e$ for a given query $q$ is defined in the same way as the CombMNZ method, but the scores of the candidates are divided by the number of ranking features that contribute to the retrieval of the candidate, instead of being multiplied, as shown in Equation~\ref{eq:CombANZ}. CombANZ gives more weight to the candidates that are relevant to the query but are not returned by many of the systems.

\begin{equation}
\text{CombANZ(e,q) = }\frac{ \text{CombSUM(e, q)}} {r_e}
\label{eq:CombANZ}
\end{equation}

The  {\bf Borda Fuse} positional method was originally proposed by~\citeasnoun{borda81} in the scope of social voting theory. This method determines the highest ranked expert by assigning to each individual candidate a certain number of votes. This number corresponds to its position in a ranked list that is given by each feature. Generally speaking, if a given candidate $e_j$ appears in the top of the ranked list, then one assigns to him $n$ votes, where $n$ is the number of experts in the list. If the candidate appears in the second position of the ranked list, then it is assigned $n-1$ votes, and so on. The final Borda score is given by the aggregation of each of the individual scores that is obtained by the candidate for each individual feature.  

The {\bf Reciprocal Rank Fuse} positional method was originally proposed by~\citeasnoun{Voorches99trec} in the scope of Question Answering systems. The reciprocal rank fuse determines the highest ranked expert by assigning to each individual candidate a certain score that corresponds to the inverse of its position in a ranked list given by each feature. Generally speaking, if a candidate $e_j$ appears at the top of the ranked list, one assigns to him $1/1$ votes. If the candidate appears in the second position of the ranked list, then he is assigned $1/2$ votes, and so on. The final reciprocal rank score is given by the aggregation of each of the individual scores obtained by the candidate for each individual feature. 

The {\bf Condorcet Fusion} majoritarian method was originally proposed by~\citeasnoun{Montague02condorcet} in the scope of social voting theory. The Condorcet Fusion method determines the highest ranked expert by accounting for the number of times that an expert wins or ties with every other candidate in a pairwise comparison. To rank the candidate experts, we use their win and loss values. If the number of wins of an expert is higher than those for another expert, then the first expert wins. A tie in Condorcet Fusion occurs when two experts have the same number of wins. To untie them, we account for the number of losses. The expert that has the lowest number of losses wins. If the experts have exactly the same number of wins and losses (an unlikely scenario, but possible), then there is no way to untie them, and the system returns both in a random order~\cite{Bozkurt07fusion}.

To give an illustrative example of how the framework for rank aggregation works, let us assume that a user wants to know the top experts in \emph{Information Retrieval}. The first step of our system is to retrieve all of the authors that have the query topics in their publication's titles or abstracts. Each set of features is then responsible for detecting different types of information in those documents. The textual features will collect information, such as the term frequency (Section~\ref{features} will detail these features). Profile features, on the other hand, will collect the total publication record of the candidate. Citation features must collect information such as the number of citations of the candidate's work and the number of co-authors. For each feature, a score will be computed and assigned to the author. These features will represent the author's knowledge of the query topics. Then, according to the data fusion method that is used, all of the individual scores of each feature will be combined into a single value. For example, if one uses the CombSUM data fusion method, then the scores of each feature will be summed. The authors are then ranked by the resulting summation score.

\section{Features for Estimating Author Expertise}\label{features}

The considered set of features for estimating the degree of expertise of a person toward a given query can be divided into three groups, namely, the textual features, profile features and citation graph features. The textual features are similar to those used in standard text retrieval systems and also in previous learning to rank experiments (e.g., TF-IDF and BM25 scores). The profile information features correspond to importance estimates for the authors, which are derived from their profile information (e.g., the number of papers published). Finally, the graph features correspond to importance and relevance estimates that are computed from citation counts and citation graphs.

\subsection{Textual Similarity Features}

Similar to previous expert finding studies that are based on document-centric approaches, we also use textual similarities between the query and the contents of the documents to build estimates of expertise. In the domain of academic digital libraries, the associations between documents and experts can easily be obtained from the authorship information that is associated with the publications. For each topic-expert pair, we used the Okapi BM25 document-scoring function to compute the textual similarity features. Okapi BM25 is a state-of-the-art IR ranking mechanism that is composed of several simpler scoring functions with different parameters and components (e.g., term frequency and inverse document frequency). It can be computed through the formula shown in Equation~\ref{eq:bm25}, where $Terms(q)$ represents the set of terms from query {\it q}, $Docs(a)$ is the set of documents that have {\it a} as an author, ${\it Freq(i,d_j)}$ is the number of occurrences of term $i$ in document $d_j$, $|d_j|$ is the number of terms in document $d_j$, and $\mathcal{A}$ is the average length of the documents in the collection. The values given to the parameters $k_1$ and $b$ were 1.2 and 0.75, respectively. Most of the previous IR experiments use these default values for the $k_1$ and $b$ parameters.

\begin{equation} \label{eq:bm25}
\begin{split}
{ \text{BM25}(q,a)} = \sum_{j \in Docs(a)}\sum_{i \in Terms(q)}\log \left( \frac{N-\text{Freq}(i, d_j)+0.5}{\text{Freq}(i, d_j)+0.5} \right) \times\\ \frac{(k_1+1) \times \frac{\text{Freq}({i,d_j})}{|d_j|}} {\frac{\text{Freq}({i,d_j})}{|d_j|} + k_1 \times (1 - b + b \times \frac{|d_j|}{\mathcal{A}})}
\end{split}
\end{equation} 

We also experimented with other textual features that are commonly used in ad-hoc IR systems, such as {\it Term Frequency} and {\it Inverse Document Frequency}.

Term Frequency (TF) corresponds to the number of times that each individual term in the query occurs in all of the documents that are associated with the author. Equation~\ref{eq:tf} describes the TF formula.

\begin{equation}
\text{TF}_{q,a} =  \sum_{j \in Docs(a)} \sum_{i \in Terms(q)} \frac{\text{Freq}({i,d_j})}{|d_j|}
\label{eq:tf}
\end{equation}

The Inverse Document Frequency (IDF) is the sum of the values for the inverse document frequency of each query term and is given by Equation~\ref{eq:idf}. In this formula, $|D|$ is the size of the document collection, and $f_{i, D}$ corresponds to the number of documents in the collection in which the $i_{th}$ query term occurs.
		
\begin{equation}
	\label{eq:idf}
	\text{IDF}_q = \sum_{i \in Terms(q)} \log \frac{|D|}{f_{i, D}}
\end{equation}

Other features that we used correspond to the number of unique authors that are associated with documents that contain the query topics, the range of years since the first and last publications of the author containing the query terms, and the sum of the document lengths, in terms of the number of words, for all of the publications that are associated with the author.

In the computation of these textual features, we considered two different textual streams from the documents, namely (i) a stream that is composed of the titles, and (ii) a stream that uses the abstracts of the articles.

\subsection{Profile Information Features}

We also considered a set of profile features that are related to the amount of published materials associated with authors, generally assuming that the expert authors are likely to be more productive. Most of the features that are based on profile information are query independent, meaning that they have the same value for different queries. The considered set of profile features is based on the temporal interval between the first and the last publications, the average number of papers and articles per year, and the number of publications in conferences and in journals with and without the query topics in their contents.

\subsection{Features Based on Citation Patterns}

Scientific impact metrics computed over scholarly networks that encode citation information can offer effective approaches for estimating the importance of the contributions of specific publications, publication venues, or individual authors. Thus, we have considered a set of features that estimate expertise based on citation information. The considered features are divided into three subsets, namely (i) citation counts, (ii) academic indexes and (iii) graph centrality. With regard to citation counts, we used the total, the average and the maximum number of citations of the papers that contain the query topics, the average number of citations per year of the papers that are associated with an author and the total number of unique collaborators that worked with an author. With regard to academic impact indexes, we used the following features:

\begin{itemize}
	
\item {\bf Hirsch index} of the author and of the author's institution, measuring both the scientific productivity and the scientific impact of the author or his institution~\cite{hirsch05index}. A given author or institution has a Hirsch index of $h$ if $h$ of his $N_p$ papers have at least $h$ citations each and the other $(N_p - h)$ papers have at most $h$ citations each. Authors who have a high Hirsch index or authors who are associated with institutions that have a high Hirsch index are more likely to be considered experts.

\item {\bf Hirsch index considering query topics} of the author, enabling the measurement of the scientific impact of the author in the field that is characterised by the query topic. An author has an {\it h} index of {\it h} if {\it h} of his $N_p$ papers that contain the query terms have at least {\it h} citations each, and the other $(Np-h)$ papers have at most {\it h} citations each.

\item {\bf Contemporary Hirsch index} of the author, which adds an age-related weighting to each cited article, giving less weight to older articles~\cite{Sidiropoulos07index}. A researcher has a contemporary Hirsch index $h^c$ if $h^c$ of his $N_p$ articles have a score of $S^c(i) >= h^c$ each, and the remaining $(N_p - h^c)$ articles have a score of $S^c(i) <= h^c$. For an article $i$, the score $S^c(i)$ is:

\begin{equation}
S^c(i) = \\
  \gamma * (\text{Y}(now) - \text{Y}(i) + 1)^{-\delta} * |\text{CitationsTo}(i)|
\end{equation}

In this formula, $Y(i)$ refers to the year of publication for article $i$. The $\gamma$ and $\delta$ parameters were set to $4$ and $1$, respectively, which means that the citations for an article that was published during the current year is accounted for as 4 times, the citations for an article that was published 4 years ago is accounted for as only one time, and the citations for an article that was published 6 years ago is accounted for as $4/6$ times, and so on.

\item {\bf Trend Hirsch index}~\cite{Sidiropoulos07index} for the author, which assigns to each citation an exponentially decaying weight according to the age of the citation, estimating the impact of a researcher's work in a specific time instance. A researcher has a trend Hirsch index $h^t$ if $h^t$ of his $N_p$ articles receive a score of $S^t(i) >= h^t$ each, and the remaining $(N_p - h^t)$ articles receive a score of $S^t(i) <= h^t$. For an article $i$, the score $S^t(i)$ is defined as shown below:

\begin{equation}
S^t(i) = \gamma * \sum_{\forall x \in C(i)} (\text{Y}(now) - \text{Y}(x) + 1)^{-\delta}
\end{equation}

Similar to the case of the contemporary Hirsch index, the $\gamma$ and $\delta$ parameters are set here to $4$ and $1$, respectively.

\item {\bf Individual Hirsch index} of the author, which is computed by dividing the value of the standard Hirsch index by the average number of authors in the articles that contribute to the Hirsch index of the author, reduces the effects of frequent co-authorship with influential authors~\cite{batista06index}.

\item The {\bf $a$-index} of the author or the author's institution, which measures the magnitude of the most influential articles. For an author or an institution that has a Hirsch index of $h$ and that has a total of $N_{c,tot}$ citations toward his papers, we say that he has an $a$-index of $a = N_{c,tot} / h^2$.

\item The {\bf $g$-index} of the author or his institution, which also quantifies scientific productivity that is based on the publication record~\cite{Egghe06index}. Given a set of articles that are associated with an author or an institution and that are ranked in decreasing order of the number of citations that they received, the g-index is the (unique) largest number such that the top $g$ articles received on average at least $g^2$ citations.

\item The {\bf $e$-index} of the author~\cite{ZhangEIndex}, which represents the excess amount of citations of an author. The motivation behind this index is that we can complement the $h$-index by accounting for the excess citations that are ignored by the $h$-index. The $e$-index is given by the formula shown in Equation~\ref{eq:e-index}:

\begin{equation}
e^2 = \sum_{j=1}^{h} cit_j-h^2 \implies e = \sqrt{\sum_{j=1}^{h} cit_j-h^2}
\label{eq:e-index}
\end{equation}

In Equation~\ref{eq:e-index}, $cit_j$ are the citations that are received by the $j_{th}$ paper and $h$ is the $h$-index.

\end{itemize}

In addition to these features and following the ideas of~\citeasnoun{Chen07Finding}, we have also considered a set of graph-centrality features that estimate the influence of individual authors using PageRank, which is a well-known graph linkage analysis algorithm that was introduced by the Google search engine~\cite{Brin99PageRank}. PageRank assigns a numerical weighting to each element of a linked set of objects (e.g., hyperlinked Web documents or articles in a citation network) with the purpose of measuring its relative importance within the set. The PageRank value of a node is defined recursively and depends on the number and PageRank scores of all of the other nodes that link to it. A node that is linked to many nodes with high PageRank scores receives a high score itself. 

Formally, given a graph with $N$ authors as nodes connected between each other through citation links, the PageRank of an author $A$, $PR(A)$, is defined by Equation~\ref{eq:PageRank}.

\begin{equation}
PR(A) = \frac{ (1 - d) }{ N } + d \sum_{j \in inLinks_A} \frac{\text{PR}(j)}{ OutLinks_j } 
\label{eq:PageRank}
\end{equation}

In Equation~\ref{eq:PageRank}, the sum is over all authors $j$ that cite author $A$. The term $OutLinks_j$ corresponds to all citations made by an author $j$, and the term $1 - d$ corresponds to the dumping factor, which can be seen as a decay factor. Under a web search scenario, it represents the probability that a user will stop clicking links and jumps to another random page. Under the expert finding scenario, the dumping factor can be seen as an interest over a different author, instead of the search process only being interested in authors that are cited. The parameter $d$ was set to 0.85, because most of the implementations in the literature use this value~\cite{Brin99PageRank}.

The PageRank-based features that we considered correspond to the sum and average of the PageRank values that are associated with the papers of the author that contain the query terms, which are computed over a directed graph that represents citations between papers. Authors who published papers with high PageRank scores are more likely to be considered experts.

\section{Experimental Validation}~\label{sec:validation}

This section describes the validation of the main hypothesis behind this work, which states that either learning to rank approaches or rank aggregation methods can combine multiple estimators of expertise in a principled way, in this way improving over the current state-of-the-art expert retrieval system. 

\subsection{Experimental Setup} \label{sec:exp}

The validation of the proposed approaches requires a sufficiently large repository of textual content that describes the expertise of the individuals. In this work, we used a dataset for evaluating expert searches in the Computer Science domain, which corresponds to an enriched version of the DBLP\footnote{\url{http://www.arnetminer.org/citation}} database that was made available through the Arnetminer project.

DBLP data have been used in several previous experiments that involve citation analysis~\cite{Sidiropoulos05citation,Sidiropoulos06generalized} and expert search~\cite{Deng08formal,deng11enhanced}. DBLP is a large dataset that covers both journal and conference publications for the computer science domain, in which substantial effort has been invested in the problem of author identity resolution (i.e., resolving to the same person possibly with different names). This dataset contains only the publications' title and, in some papers, the abstract. The contents of the full paper are not available. Table~\ref{t1} provides a statistical characterisation of the DBLP dataset.

The main reason for using the DBLP dataset is the fact that, to the best of our knowledge, it is the only dataset for expert finding in academic publications that provides relevance judgments for each query-expert pair. Therefore, it was the only dataset that is publicly available that enabled the exploration of supervised techniques. Moreover, this dataset is the most often used in the literature for finding academic experts\cite{Yang09bole,Deng08formal,deng11enhanced}. We note that our approach could be extended to any dataset of academic publications, as long as the dataset provides information about the publication's titles and abstracts, their respective authors and the references in the publications.

The DBLP dataset contains many types of publications (e.g., books, journals, conference proceedings, etc). In our paper, we considered only journal papers and conference papers, because they are the publications that researchers have more access to. For this reason, in Table~\ref{t1}, the number of total publications is different from the sum of conference and journal publications.

\begin{table*}[ht]
\begin{center}
\begin{tabular}{l c}
  
  Characterisation															& Statistics								\\
  \hline
  Total Authors 																& ~~~~1 033 050~~~~		\\
  Total Publications 														& ~~~~1 632 440~~~~ 		\\
  Total Publications containing Abstract										& ~~~~653 514~~~~ 			\\
  Total Papers Published in Conferences~~~~~~~~~~~~~~~~~~~~~~~~~~~~~~~~~~~	& ~~~~606 953~~~~			\\
  Total Papers Published in Journals 										& ~~~~436 065~~~~ 			\\
  Total Number of Citations Links 											& ~~~~2 327 450~~~~ 		\\
  \hline
\end{tabular}
\end{center}
\scriptsize
\caption{Statistical characterization of the DBLP dataset used in our experiments}
\label{t1}
\end{table*}

To validate the different approaches that have been proposed in this work, we required a set of queries that have the corresponding author relevance judgments. We used the relevance judgments that were provided by Arnetminer,\footnote{\url{http://arnetminer.org/lab-datasets/expertfinding/}} that have already been used in other expert finding experiments~\cite{Yang09bole,deng11enhanced}. The Arnetminer dataset comprises a set of 13 query topics from the Computer Science domain, and it was built by collecting people from the program committees of important conferences that are related to the query topics. Table~\ref{judgements} shows the distribution of experts that are associated with each topic, as provided by Arnetminer.

\begin{table*}[ht]
\begin{center}
\begin{tabular}{l c l c}

~~{\bf Query Topics}								& ~~{\bf Rel. Authors}		&~~{\bf Query Topics} 						&~~{\bf Rel. Authors}	\\
\hline

 ~~Boosting (B)											& ~~46								&	~~Natural Language (NL)					& ~~41							\\
 ~~Computer Vision (CV)							&	~~176							&	~~Neural Networks	 (NN)					& ~~103						\\
 ~~Cryptography (C)									&	~~148							& ~~Ontology				 (O)					& ~~47							\\
 ~~Data Mining (DM)								&	~~318							& ~~Planning				 (P)					& ~~23							\\
 ~~Information Extraction (IE)					&	~~20							& ~~Semantic Web		 (SW)					& ~~326						\\
 ~~Intelligent Agents	(IA)						&	~~30							& ~~Support Vector Machines (SVM)	& ~~85							\\
 ~~Machine Learning	(ML)						& ~~34								& 															&									\\
\hline
\end{tabular}
\end{center}
\caption{Characterisation of the Arnetminer dataset of Computer Science experts.}
\label{judgements}
\end{table*}

With respect to the learning to rank framework, we used existing learning to rank implementations that are available on the RankLib\footnote{\url{http://www.cs.umass.edu/~vdang/ranklib.html}} software package that was developed by Van Dang as well as on the SVMrank\footnote{\url{http://www.cs.cornell.edu/people/tj/svm_light/svm_rank.html}} implementation by~\citeasnoun{Joachims06rank}, on the SVMmap\footnote{\url{http://projects.yisongyue.com/svmmap/}} implementation by~\citeasnoun{Yue07Support} and in the Additive Groves algorithm implemented by~\citeasnoun{Sorokina07groves}. All of the considered algorithms have already been described in Section~\ref{sec:l2r} of this work.

To train and validate the different learning to rank algorithms, because the Arnetminer dataset contains only relevant people for all 13 query topics, we had to complement this dataset with negative relevant judgments (i.e., adding unimportant authors for each of the query topics). Because the supervised machine learning methods that we use in this paper have difficulty in addressing unbalanced datasets, we balanced our data by adding non-relevant authors in equal numbers to the relevant authors, prior to model training. We also added non-relevant cases that were both easy and hard to classify. The general idea is to select non-relevant authors by random sampling (half of them) and also to select authors who score high by using the considered features (e.g., BM25). Because in learning to rank algorithms, the goal is to learn a ranking order rather than performing a classification over instances of classes, the test set that was built is a fair approach to verify whether the learning to rank algorithms, that are proposed in the literature, are suitable for the task of expert finding.

The test collection was used in a $k$-fold cross-validation methodology, in which we used 4-folds. In this method, our data were randomly partitioned into $k$ equal-size sub-samples. By equal size, we mean that the number of different classes (queries) for classification is the same in every sub-sample. Of these $k$ sub-samples, a single subsample was retained as the validation data to test the model, and the remaining $k - 1$ sub-samples were used as training data. The cross-validation was then repeated $k$ times, where each of the k subsamples were used exactly once as validation data. The results of the $k$ folds were then averaged to produce a single estimation. In the end, each fold contained 9 queries to train and 4 queries to test.

The parameters were determined by using a grid search approach (see Section~\ref{sec:parameter_est}). Table~\ref{tab:parameters} presents the optimal parameters that were found.

\begin{table}[ht]%
\resizebox{\columnwidth}{!} {
\begin{tabular}{l c c c c c c}
Parameters						& AdaRank		& Coordinate Ascent		& RankBoost	& RankNet	& SVMrank	& SVMmap\\
\hline
\# Iterations~					&	400			&		100				& 	 300 	& 30			& -			&	-	\\
\# Threshold Candidates ~		&		-		&		-				&	 40		&	-		& -			& 	-	\\
\# Random Restarts~				&		-		&		5				&	   -		&	-		& -			& 	-	\\
\# Nodes in Hidden Layer~ 		&		-		&		-				&		-	& 50			& -			&	-	\\
~C value (SVM only)				& 		-		& -						& - 			& - 			& 900		&	8		\\
~Gamma value (SVM only)			& 		-		& -						& - 			& - 			& 	-		&	0.125		\\
\hline

\end{tabular}
}
\caption{Parameters found with the grid search approach. The "-" symbol means that the algorithm does not have the parameter. For instance, AdaRank only has one parameter which is the number of iterations. The other parameters do not apply to this algorithm}
\label{tab:parameters}
\end{table} 

With regard to the rank aggregation framework, we implemented six different data fusion algorithms that were based in positional methods, majoritarian methods and score aggregation. The score aggregation algorithms that were developed were CombSUM, CombMNZ and CombANZ. The positional algorithms were the Borda Fuse and Reciprocal Rank Fuse. The majoritarian algorithm was Condorcet Fusion. All of these algorithms have been described in Section~\ref{sec:rankaggregation} of this work.

To validate the different rank aggregation algorithms, we again had to complement the Arnetminer dataset with negative relevance judgements (i.e., adding unimportant authors for each of the query topics). Because we are interested in deriving a ranking ordering, rather than a classification, the non-expert candidates were obtained in the following way: we retrieved the top authors that were not marked as relevant from the database for each query topic according to the BM25 metric. The Arnetminer instances together with the non-relevant candidates that were collected constituted a total of 350 instances for each query topic. 

To measure the quality of the results, we used two different performance metrics, namely, the Precision at $k$ (P@k) and the Mean Average Precision (MAP).

The Precision at rank $k$ is used when a user wishes to look only at the first $k$ retrieved domain experts. The precision is calculated at that rank position through Equation~\ref{eq:PrecisionRank}.

\begin{equation}
\text{P@k}=\frac{r\left(k\right)}{k}
\label{eq:PrecisionRank}
\end{equation}

In the formula, $r(k)$ is the number of relevant authors that are retrieved in the top {\it k} positions. $P@k$ considers only the top-ranking experts to be relevant and computes the fraction of such experts in the top $k$ elements of the ranked list.

The Mean of the Average Precision over the test queries is defined as the mean over the precision scores for all of the retrieved relevant experts. For each query $r$, the Average Precision (AP) is given by:

\begin{equation}
\text{AP}[r] = \frac{\sum_{k=1}^n \text{P@k}[r] \times I\{ g_{r_k} = \max(g) \}}{\sum_{k=1}^n I\{ g_{r_k} = \max(g) \}}  
\label{eq:AP}
\end{equation}

In Equation~\ref{eq:AP}, $n$ is the number of experts that are associated with query $q$, and $g_{rk}$ is the relevance grade for author $k$ in relation to query $r$. In the case of our datasets, $\max(g) = 1$ (i.e., we have 2 different grades for relevance, 0 or 1).

The Normalised Discount Cumulative Gain (NDCG) emphasises the fact that the highly relevant experts should appear on top of the ranked list. This metric is given by Equation~\ref{eq:ndcg}, where $Z_k$ is a normalisation factor that corresponds to the maximum score that could be obtained when looking at the top experts, and $rel_i$ is the relevance score that is assigned to expert $i$.

\begin{equation} \label{eq:ndcg}
\text{NDCG}[r] = Z_k \sum_{i = 1}^{k} \frac{2^{rel_i} - 1}{\log_2(1+i)}
\end{equation}

We also performed statistical significance tests over the results by using an implementation\footnote{\url{http://www.mansci.uwaterloo.ca/~msmucker/software/paired-randomization-test-v2.pl}} of the two-sided randomisation test~\cite{Smucker07StatSig}.

\subsection{Experimental Results}~\label{sec:results}

This section presents the results that were obtained in both of the frameworks that were tested in this work, namely the supervised learning to rank framework and the rank aggregation framework for expert finding.

\subsubsection{The Learning to Rank Framework}

In this paper, we argue that the use of supervised learning to rank algorithms is a sound approach for the expert finding task, effectively combining a large pool of estimators that characterise the knowledge of an expert.

The goal of a learning to rank framework is to combine features in an optimal way. In our work, we combine them by using different learning to rank algorithms, and we test them in two different ways: (1) by determining the best algorithm for the expert finding task in academic publications (Table~\ref{t2}), and (2) by comparing different groups of features to understand which groups achieve better results and are more relevant to discriminating experts (Table~\ref{t3}).

Table~\ref{t2} presents the results that were obtained over the DBLP dataset. These results show that the pointwise approach Additive Groves outperformed all of the other pairwise and listwise learning to rank algorithms, in terms of MAP, and therefore provided a better ranking procedure than all of the other approaches that were tested. On the other hand, the listwise SVMmap algorithm, with an RBF kernel, performed almost as good as the Additive Groves method. This finding makes sense because the goal of SVMmap is to optimise the Mean Average Precision scores. In fact, SVMmap outperformed Additive Groves when ranking the top 5 experts (P@5), which shows that this listwise method can also be successfully used in the context of expert finding.

Table~\ref{t2} also shows competitive results for the pairwise approach of SVMrank with a linear kernel and for RankBoost. This finding leads to the conclusion that pairwise learning to rank algorithms that are based on the formalisms of support vector machines and on the boosting framework can also be applied successfully in the task of expert finding. 

With respect to the worst results that were obtained, Table~\ref{t2} shows that AdaRank and RankNet were not as successful as the other algorithms. For RankNet, an easy explanation can be found in the use of gradient descent in the backpropagation algorithm. Gradient descent is an optimisation algorithm that attempts to find the minimum of a function by taking steps that are proportional to the negative gradient of the function. However, this method has the problem of not being able to find a global minimum because it can get stuck in a local minimum and, therefore, a good optimisation might never be achieved. In addition, the cost function that is used in RankNet is general and does not correspond to any specific information-retrieval metric. For the case of AdaRank, its greedy feature selection method did not perform very well in this experiment. Although AdaRank and RankBoost are based on the same paradigm, i.e., boosting, they use it in very different ways. For example, RankBoost contains an extra parameter, which is the number of threshold candidates. AdaRank, on the other hand, makes a greedy linear combination of weak rankers. RankBoost not only makes a linear combination of weak rankers, but also considers whether an expert who is returned by a query is above a specific threshold (this approach decreases the weight of the weak rankers of correctly ranked experts and consequently builds more accurate ranking functions). 

Table~\ref{t2} presents the results of the baseline models that were proposed by~\citeasnoun{Balog06FormalModels}, namely the candidate-based Model 1 and the document-based Model 2. To make the comparison fair, we used the code that was made publicly available by Balog\footnote{\url{http://code.google.com/p/ears/}}. Experiments revealed that Model 1 and Model 2 have similar performances on such an academic dataset, but they achieved a lower performance when compared to all of the algorithms that were tested with the supervised learning approach. In Model 1, when an author publishes a paper that contains a set of words that exactly match the query topics, the author achieves a very high score. In addition, because we are addressing very large datasets, there are many authors in such a situation and, consequently, the top-ranked authors are dominated by non-experts while the real experts are ranked lower. Thus, more general papers on a subject can actually contain the topic in the title and abstract compared with papers that are written by experts, who tend to jump directly into the details. In Model 2, because we only include the publications' titles and some abstracts, the query topics might not occur very often in publications that are associated with expert authors. In such a model, the final ranking of a candidate is given by aggregating the scores that he achieved in each publication. If the document's abstract or title does not contain the query topic or has a small connection to the query topic, then the candidate will receive a lower score in the final ranked list.

Finally, Table~\ref{t2} shows that all of the algorithms tested in this supervised framework outperformed the baseline ranking function BM25 as well as Models 1 and 2 from~\citeasnoun{Balog06FormalModels}. Table~\ref{t2_pvals} shows the p-values that were obtained for statistical significance tests over these results, using a two-sided randomisation test, to compare the different approaches against the Additive Groves model.

\begin{table*}[ht]
\begin{center}
\resizebox{\columnwidth}{!} {
\begin{tabular}{ l c c c c c c }

L2R Algorithms 	&~~~~P@5~~~~ 	& ~~~~P@10~~~~ 	& ~~~~P@15~~~~ 	& ~~~~P@20~~~~ 	& ~~~~MAP~~~~ 	& ~~~~NDCG~~~~ 	\\
\hline
~~AdaRank	        & 0.667			& 0.683			& 0.674			& 0.681			& 0.648			& 0.885		  			\\
~~Coordinate Ascent & 0.925			& 0.873			& 0.841			& 0.825			& 0.758			& 0.936		  			\\
~~RankNet	        & 0.704			& 0.719			& 0.688			& 0.676			& 0.653			& 0.873		  			\\
~~RankBoost         & 0.838			& 0.879			& 0.832			& 0.815			& 0.784			& 0.940		  			\\
~~Additive Groves   & 0.967			& {\bf 0.958}	& {\bf 0.947}	& {\bf 0.940}	&{\bf 0.894}		& {\bf 0.973}		  	\\
~~SVMmap		        & {\bf 0.989}	& 0.921			& 0.897			& 0.892			& 0.870			& 0.970		  			\\
~~SVMrank      	    	& 0.954			& 0.913			& 0.882			& 0.883			& 0.831			& 0.956		  			\\
\hline
~~BM25 (baseline)	& 0.692			& 0.577			& 0.503			& 0.477			& 0.542			& 0.838 		\\
~~Balog's Model 1~\cite{Balog06FormalModels}	& 0.077	 & 0.085 & 0.108	& 0.135			& 0.221			& 0.632		\\
~~Balog's Model 2~\cite{Balog06FormalModels}	& 0.077	 & 0.100 & 0.097	& 0.131			& 0.214			& 0.628		\\		
\hline
\end{tabular}
}
\end{center}
\caption{Results for the various learning to rank algorithms that were tested.}
\label{t2} 
\end{table*}

\begin{table}
\resizebox{\columnwidth}{!} {
\begin{tabular}{l c c c c c c}
Algorithms					& P@5			& P@10			& P@15		 & P@20			& MAP			& NDCG \\
\hline
Add. Groves vs AdaRank		& 0.00200* 		& 0.00049*		& 0.00051*  & 0.00021* 	& 0.00038*		& 0.00024*	\\
Add. Groves vs Coord. Ascent	& 0.49894~		& 0.03075*		& 0.02347*	 & 0.00797*		& 0.00327*		& 0.00701* 	\\
Add. Groves vs RankNet		& 0.01578*		& 0.00215*		& 0.00052*	 & 0.00021*		& 0.00038*		& 0.00024*	\\
Add. Groves vs RankBoost		& 0.49717~		& 0.08504~		& 0.03020*	 & 0.00639*		& 0.00941*		& 0.03638* 	\\
Add. Groves vs SVMmap		& 1.00000~		& 0.32419~		& 0.21003~	 & 0.13053~		& 0.33701~		& 0.67007~	\\
Add. Groves vs SVMrank		& 1.00000~		& 0.19688~		& 0.10640~	 & 0.11147~		& 0.08929~		& 0.27928~	\\
Add. Groves vs BM25			& 0.03448*		& 0.00137*		& 0.00052*	 & 0.00021*		& 0.00065*		& 0.00083*	\\
Add. Groves vs Balog's Model 1~\cite{Balog06FormalModels}	& 0.00021*		& 0.00023*		& 0.00022*		& 0.00021*		& 0.00038*			& 0.00024*	\\
Add. Groves vs Balog's Model 2~\cite{Balog06FormalModels}	& 0.00021*			& 0.00023*		& 0.00022*		& 0.00021*		& 0.00038*			& 0.00024*	\\
\hline
\end{tabular}
}
\caption{P-Values obtained during significance tests. The * indicates that the improvement obtained using the Additive Groves algorithm is statistically significant, for a confidence interval of 95\%.}
\label{t2_pvals}
\end{table}

In a separate experiment, we attempted to measure the impact of the different types of ranking features on the quality of the results. Using the best performing learning to rank algorithm, namely the Additive Groves method, we measured the results that were obtained by ranking models that considered (i) only the textual similarity features, (ii) only the profile features, (iii) only the graph features, (iv) textual similarity and profile features, (v) textual similarity and graph features and (vi) profile and graph features. Table~\ref{t3} shows the results that were obtained, and Table~\ref{t3_pvals} shows the p-values for the statistical significance tests, using a two-sided randomisation test.

\begin{table*}[ht]
\begin{center}
\resizebox{\columnwidth}{!} {
\begin{tabular}{ l c c c c c c}
    Sets of Features                 	&~~~~P@5~~~~ 		& ~~~~P@10~~~~ 	& ~~~~P@15~~~~ 	& ~~~~P@20~~~~ 	& ~~~~MAP~~~~  & ~~~~NDCG~~~~\\
\hline
Text Similarity + Profile + Graph ~~ 	& {\bf 0.967} 		& {\bf 0.958}	& {\bf 0.947}	& {\bf 0.940}	& {\bf 0.894} & {\bf 0.973}\\
Text Similarity + Profile				& {\bf 0.967}		& 0.944			& 0.936			& 0.917			& 0.871		   & 0.963	  	\\
Text Similarity + Graph					& 0.933				& 0.944			& 0.936			& 0.924			& 0.883		   & 0.969	  	\\
Profile + Graph							& 0.921				& 0.875			& 0.863			& 0.850			& 0.824		   & 0.946	  	\\
Text Similarity        					& 0.933				& 0.919			& 0.913			& 0.901			& 0.866		   & 0.965		  \\
Profile       							& {\bf 0.967} 		& 0.910			& 0.915			& 0.913			& 0.873		   & 0.967		  \\
Graph        							& 0.892				& 0.917			& 0.893			& 0.895			& 0.853		   & 0.957\\
\hline
\end{tabular}
}
\end{center}
\caption{The results obtained with different sets of features.}
\label{t3}
\end{table*}

\begin{table}
\resizebox{\columnwidth}{!} {
\begin{tabular}{l c c c c c c}
Sets of Features								& P@5				& P@10			& P@15		 	& P@20			& MAP				& NDCG \\
\hline
All Features vs Text Similarity + Profile	& 1.00000~		& 0.62337~		& 0.74859~		& 0.38971~		& 0.02632*			& 0.02494* \\
All Features vs Text Similarity + Graph		& 0.49951~		& 0.50039~		& 0.37410~		& 0.24767~		& 0.02207*			& 0.02850* \\
All Features vs Profile + Graph				& 0.50069~		& 0.12367~		& 0.05410~		& 0.03009*		& 0.01128*			& 0.00100*	\\
All Features vs Text Similarity				& 0.49951~		& 0.12539~		& 0.12440~		& 0.03976*		& 0.08140~			& 0.05283~ 	\\
All Features vs Profile						& 1.00000~		& 0.31476~		& 0.50108~		& 0.73553~		& 0.53595~			& 0.56721~	\\
All Features vs Graph						& 0.18783~		& 0.24849~		& 0.06149~		& 0.10932~		& 0.02725*			& 0.57462~	\\
\hline
\end{tabular}
}
\caption{P-Values obtained during significance tests. The * indicates that the improvement obtained using the Additive Groves algorithm using the entire set of features is statistically significant, for a confidence interval of 95\%.}
\label{t3_pvals}
\end{table}

Table~\ref{t3} shows that the set that has the combination of all the features has the best results. The results also show that the combination of profile features that includes graph features has the poorest results. This finding means that the presence of the query topics in the author's publications, specifically in the titles and abstracts, is crucial to determining whether some authors are experts or not, and indeed, the information that is provided by textual evidence can help in expertise retrieval. 

\subsubsection{The Rank Aggregation Framework}

In this paper, we also argue that rank aggregation methods, which are based on data fusion techniques, can provide significant advantages over the representative generative probabilistic models that are proposed in the expert finding literature. We used existing state-of-the-art algorithms to build an expert finding framework that enables a combination of a large pool of expertise estimates. 

Table~\ref{t4} presents the obtained results on the DBLP dataset. The CombMNZ rank aggregation technique outperformed all of the other algorithms, in terms of MAP, which shows that this rank aggregation method provides a better ranking than all of the other approaches. On the other hand, the Condorcet Fusion algorithm outperformed all of the other methods in almost all of the evaluation metrics tested. In fact, the Condorcet Fusion method achieved much better results for the P@$k$ than all of the other algorithms. Both MAP and NDCG are important metrics for the evaluation of system performance. However, P@$k$ also plays an important role in the evaluation process, given that, in expert finding systems, when the user searches for experts on some topic, he is usually only interested in the top $k$ experts that are retrieved.

Table~\ref{t4} also shows that the Borda Fuse and the Reciprocal Rank Fuse algorithms had the same performance in our experiments. This finding is not surprising because these positional algorithms are very similar. The only difference between them is that Borda Fuse uses directly the positions of the candidates, whereas Reciprocal Rank Fuse uses the reciprocal rank of the positions of the candidates. In this experiment, the final ranked lists were the same, in other words, they returned the experts in the same order.

Regarding the worst results that were obtained, Table~\ref{t4} shows that CombANZ was not as successful as the other algorithms. This finding can be explained by the fact that the CombANZ algorithm divides the CombSUM scores by the number of systems that contribute to the ranking of a candidate. Because the validation data were manipulated such that all of the candidates had the presence of the query topics in their publications, CombANZ de-emphasises those candidates that appear multiple times in the different systems.

Finally, Table~\ref{t4} shows that all of the algorithms that were tested in this rank aggregation framework outperformed the baseline ranking function BM25 and the state-of-the-art approaches proposed by~\citeasnoun{Balog06FormalModels}, namely, the candidate-based Model 1 and the document-based Model 2, for the same reasons that were noted in the L2R experiment. Table~\ref{t4_pvals} shows the p-values for statistical significance tests using a two-sided randomisation test.

\begin{table*}[ht]
\begin{center}
\resizebox{\columnwidth}{!} {
\begin{tabular}{ l c c c c c c }

Data Fusion Algorithms 	&~~~~P@5~~~~ 		& ~~~~P@10~~~~ 	& ~~~~P@15~~~~ 	& ~~~~P@20~~~~ 	& ~~~~MAP~~~~ 	& ~~~~NDCG~~~~ 	\\
\hline
~~CombSUM	          	& 0.400				& 0.408			& 0.415			& 0.450			& 0.413		  	& 0.739		 	\\
~~CombMNZ  				& 0.492				& 0.477			& 0.472			& 0.512			& {\bf 0.484} 	& 0.776		  	\\
~~CombANZ	          	& 0.308				& 0.323			& 0.380			& 0.400			& 0.356			& 0.699			\\
~~Borda Fuse          	& 0.200				& 0.223			& 0.303			& 0.342			& 0.400			& 0.701		  	\\
~~Reciprocal Rank Fuse	& 0.200				& 0.223			& 0.303			& 0.342			& 0.400			& 0.701		  	\\
~~Condorcet Fusion      & {\bf 0.708}		& {\bf 0.608}	& {\bf 0.564}	& {\bf 0.515}	& 0.438			& {\bf 0.798} 				\\
\hline
~~BM25 (baseline)       & 0.492				& 0.431			& 0.385			& 0.339			& 0.326			& 0.709		  	\\
~~Balog's Model 1~\cite{Balog06FormalModels}	& 0.077    		& 0.085			& 0.092			& 0.104			& 0.121	 & 0.544	\\
~~Balog's Model 2~\cite{Balog06FormalModels}	& 0.123			& 0.092			& 0.103			& 0.092			& 0.121	 & 0.542	 \\
\hline
\end{tabular}
}
\end{center}
\caption{Results for the various data fusion algorithms that were tested.}
\label{t4} 
\end{table*}

\begin{table}
\resizebox{\columnwidth}{!} {
\begin{tabular}{l c c c c c c}
Algorithms									& P@5			& P@10			& P@15		 & P@20			& MAP			& NDCG \\
\hline
Cond. Fusion vs CombSUM						& 0.00187* 		& 0.00584*		& 0.01272*  	& 0.10114~ 	& 0.30943~		& 0.04779*	\\
Cond. Fusion vs CombMNZ						& 0.00997*		& 0.02306*		& 0.04293*	 & 0.40498~		& 0.62124~		& 0.38286~ 	\\
Cond. Fusion vs CombANZ						& 0.00094*		& 0.00102*		& 0.00621*	 & 0.03273*		& 0.02431*		& 0.00024*	\\
Cond. Fusion vs Borda Fuse					& 0.00094*		& 0.00102*		& 0.00420*	 & 0.00830*		& 0.20045~		& 0.02883* 	\\
Cond. Fusion vs R. Rank Fuse					& 0.00094*		& 0.00102*		& 0.00420*	 & 0.00830*		& 0.20045~		& 0.02883*	\\
Cond. Fusion vs BM25							& 0.00181*		& 0.00096*		& 0.00084*	 & 0.00064*		& 0.00038*		& 0.00024*	\\
Cond. Fusion vs Balog's Model 1~\cite{Balog06FormalModels} & 0.00021*		& 0.00023*		& 0.00022*	 & 0.00021*	 & 0.00038*	& 0.00024*	\\
Cond. Fusion vs Balog's Model 2~\cite{Balog06FormalModels} & 0.00021*		& 0.00023*		& 0.00022*	 & 0.00021*	 & 0.00038*	& 0.00024*	\\
\hline
\end{tabular}
}
\caption{P-Values obtained during significance tests. The * indicates that the improvement obtained using the Condorcet Fusion algorithm is statistically significant, for a confidence interval of 95\%.}
\label{t4_pvals}
\end{table}

Again, in a separate experiment, we measured the impact of the different types of ranking features on the quality of the results. Using the Condorcet Fusion algorithm, we measured the results that were obtained by ranking models that considered (i) only the textual similarity features, (ii) only the profile features, (iii) only the graph features, (iv) the textual similarity and profile features, (v) the textual similarity and graph features and (vi) the profile and graph features. Table~\ref{t5} shows the obtained results. Table~\ref{t5_pvals} shows the statistical significance tests that were performed by using a two-sided randomisation test.

\begin{table*}[ht]
\begin{center}
\resizebox{\columnwidth}{!} {
\begin{tabular}{ l c c c c c c}
     Sets of Features               		&~~~~P@5~~~~ 	& ~~~~P@10~~~~ 	& ~~~~P@15~~~~ 	& ~~~~P@20~~~~ 	& ~~~~MAP~~~~ 	&~~~~NDCG~~~~\\
\hline
Text Similarity + Profile + Graph~~		& {\bf 0.708} 	& {\bf 0.608} 	& {\bf 0.564}	& 0.515			& 0.438			& {\bf 0.798}\\
Text Similarity + Profile				& 0.400		   	& 0.415			& 0.400			& 0.373			& 0.327			& 0.704	\\
Text Similarity + Graph					& 0.569		   	& 0.523			& 0.477			& 0.450			& 0.391			& 0.762	\\
Profile + Graph							& 0.631		   	& 0.539			& 0.482			& 0.462			& 0.417			& 0.774	\\
Text Similarity        					& 0.350		   	& 0.333			& 0.344			& 0.313			& 0.298			& 0.668	\\
Profile       							& 0.462		   	& 0.431			& 0.415			& 0.419			& 0.369			& 0.724	\\
Graph        						 	& 0.646		   	& 0.577			& 0.544			& {\bf 0.531}	& \bf{0.439}		& 0.786	\\
\hline
\end{tabular}
}
\end{center}
\caption{The results obtained with different sets of features.}
\label{t5}
\end{table*}

\begin{table}
\resizebox{\columnwidth}{!} {
\begin{tabular}{l c c c c c c}
Sets of Features						& P@5				& P@10			& P@15		 	& P@20			& MAP			& NDCG \\
\hline
All Features vs Text Similarity + Profile		& 0.00047*			& 0.00198*		& 0.00052*		& 0.00042*		& 0.00038*		& 0.00024* \\
All Features vs Text Similarity + Graph		& 0.01573*			& 0.03341*		& 0.02886*		& 0.02930*		& 0.00410*		& 0.00803* \\
All Features vs Profile + Graph			& 0.43723~			& 0.02689*		& 0.02374*		& 0.04621*		& 0.07792~		& 0.03597*	\\
All Features vs Text Similarity				& 0.00091*			& 0.00044*		& 0.00048*		& 0.00042*		& 0.00062*		& 0.00038* 	\\
All Features vs Profile					& 0.03319*			& 0.00584*		& 0.00530*		& 0.00579*		& 0.01095*		& 0.00680*	\\
All Features vs Graph					& 0.36671~			& 0.22709~		& 0.35898~		& 0.99413~		& 0.62449~		& 0.41703~	\\

\hline
\end{tabular}
}
\caption{P-Values obtained during significance tests. The * indicates that the improvement obtained using the Condorcet Fusion algorithm using the entire set of features is statistically significant, for a confidence interval of 95 \%.}
\label{t5_pvals}
\end{table}

As observed, the set with the combination of all of the features had the best results. Because DBLP has rich information about citation links, one can see that the set of graph features also achieved very competitive results. The results also show that, individually, textual similarity features have the poorest results. This finding means that considering only the textual evidence provided by query topics, together with the article's titles and abstracts, might not be sufficient to determine if some authors are experts or not and that indeed the information provided by citation patterns can help in expertise retrieval through a rank aggregation framework.

\section{Conclusions and Future Work}\label{sec:conclusions}

The tests performed in this paper indicate that learning to rank approaches achieve an overall good performance in the task of expert finding, within digital libraries of academic publications.

The various learning algorithms that were tested achieved significantly different results from one another, which leads us to conclude that some of the algorithms are more suitable for this task than others. In addition, we experimentally demonstrated that the Additive Groves pointwise approach and the SVMmap listwise approach outperformed the other algorithms. These results were quite interesting, because pointwise approaches do not consider the order of the experts and, therefore, worse results were expected for the Additive Groves approach. However, one must consider that the Additive Groves algorithm is very robust in such a way that at each iteration, it always trains a new and more accurate regression tree for the experts that were misclassified by the previous iteration and then merges all of the learned trees into a single tree. In addition, this algorithm was amongst the top 5 best performing algorithms in the Yahoo! Learning to Rank Competition~\cite{Chapelle11}. This finding implies that pointwise approaches can also be very effective, performing similarly or even better than listwise approaches. 

Regarding our rank aggregation framework, the results showed that rank aggregation approaches also provide reasonable results for the task of expert finding in digital libraries, because they outperformed some of the state-of-the-art approaches, in terms of MAP. In our experiments, CombMNZ and the Condorcet Fusion algorithms achieved the best results.

The effectiveness of the learning to rank and rank aggregation frameworks directly depends on the quality of the features that they use. As a result, in this work, both frameworks achieved very good results, always outperforming state-of-the-art approaches. We can argue that the features that are proposed can provide accurate information and discriminate the expertise levels of the candidates, enabling the retrieval of a reliable and accurate ranked list of experts. When comparing all of the different sets of features, we concluded that a combination of all of the features (textual, profile and graph) is required to achieve the best results in both experiments.

For future work, it would be very interesting to apply the algorithms that were tested in this work in the TREC enterprise task dataset, which is very commonly used in empirical studies. For example, in the learning to rank approach for expert finding that is proposed by~\citeasnoun{Macdonald11aggr}, they achieved the best results by using the AdaRank listwise algorithm, turning their approach into one of the top contributions for the expert finding task in enterprises. However, the experiments that were performed in the scope of this article showed that AdaRank was an approach that performed poorly. We are very curious to know how the Additive Groves algorithm would perform on such a task. 

It would also be interesting to extend the features that were proposed in this work to support the various expert finding models that have been proposed in the literature. For example, we could build a supervised learning to rank system by still using the Additive Groves algorithm but with a new set of features, which would be based on other studies in the literature. For example, we could use~\possessivecite{Balog06FormalModels} candidate-based model scores (Model 1),~\possessivecite{Balog06FormalModels} document-based model scores (Model 2), and~\possessivecite{deng11enhanced} query-sensitive AuthorRank scores. Given that each of these models alone already provided a good method for ranking experts, the combination of all of them through a supervised machine-learning approach could lead to even more accurate and reliable ranked lists.

\section{Acknowledgements} 

This work was supported by Funda\c{c}\~{a}o para a Ci\^{e}ncia e Tecnologia (FCT) through INESC-ID multi annual funding under project PEst-OE/EEI/LA0021/2013 and through FCT Project SMARTIS (ref. PTDC/EIA-EIA/115346/2009).

\footnotesize
\bibliographystyle{agsm}

\end{document}